\documentclass[10pt,twocolumn,english,aps,prb,showpacs,superscriptaddress]{revtex4}
\usepackage{times}
\usepackage[T1]{fontenc}
\usepackage[latin1]{inputenc}
\usepackage{amsmath}
\usepackage{babel}
\usepackage{graphics}
\usepackage{amssymb}
\usepackage{subfigure}

\makeatletter

\providecommand{\LyX}{L\kern-.1667em\lower.25em\hbox{Y}\kern-.125emX\@}

\makeatother
\begin{document}

\title{Signatures of collective local and nanoscale distortions in diffraction
experiments}

\author{Angel J. Garcia-Adeva}

\email{angarcia@lanl.gov}

\affiliation{Materials Science and Technology Division (MST-8); Mailstop G755;
Los Alamos National Laboratory; Los Alamos, NM 87545}

\author{Dylan R. Conradson}

\affiliation{Materials Science and Technology Division (MST-8); Mailstop G755;
Los Alamos National Laboratory; Los Alamos, NM 87545}

\author{Phillip Villella}

\affiliation{Materials Science and Technology Division (MST-8); Mailstop G755;
Los Alamos National Laboratory; Los Alamos, NM 87545}

\affiliation{Department of Physics, University of Colorado, Boulder, Colorado
80309}

\author{Steven Conradson}

\affiliation{Materials Science and Technology Division (MST-8); Mailstop G755;
Los Alamos National Laboratory; Los Alamos, NM 87545}

\begin{abstract}
The effects of periodic and aperiodic distortions on the structure
factor and radial distribution function of single-component lattices
are investigated. To this end, different kinds of distortions are
applied to the otherwise perfect square lattice and the corresponding
radial distribution function and structure factor for the resulting
lattices are calculated. When the applied distortions have a periodic
character, they are very easily recognized in the calculated structure
factors as new superlattice peaks. However, when the periodicity of
the distortions is suppressed the signatures of disorder only show
up as smooth and subtle features on the diffuse part of the scattering,
making it very difficult to identify the nature of the distortions
present in the lattice. The implications of these results are discussed.
\end{abstract}

\pacs{61.46.+w, 61.43.-j, 61.72.-y}

\maketitle

\section{Introduction}

Diffraction by X-rays (XRD) in the laboratory and at synchrotron facilities
and neutrons at special sources are some of the most venerable experimental
methods in Physics, having made a prominent contribution in establishing
many basic concepts in atomic and solid state physics. In particular,
our understanding of XRD is so profound that it is routinely used
for obtaining structural information about crystalline solids. In
conjunction with very powerful structural analysis algorithms, it
is a matter of hours to identify the crystallographic positions of
the atoms of an unknown crystalline material. However, diffraction
analysis relies on the well-founded paradigm of crystalline solids
as the periodic repetition of a small pattern of atoms (the unit cell).
Even though this description of solids can be considered as one of
the pillars in the construction of the modern condensed matter physics,
it is also true that the vast majority of real solids exhibit one
or another form of structural disorder. Aperiodic disorder manifests
itself in the structure factor as diffuse scattering\cite{Welberry1995,Warren1990}.
Although it is possible to extract information from the diffuse part
of the scattering, the analysis is complicated and relies strongly
on the particular structural model chosen. The possibility therefore
exists for uniqueness problems that make it very difficult to gain
reliable information about the disorder present in real solids. Thus,
conventional crystallographic analysis of diffraction patterns provides
an excellent description of the average long-range arrangements of
the coherent fraction of the atoms, without providing definitive or
even incisive information about local distortions that may be coupled
to the compelling properties of many complex materials.

To put all these facts on a more formal basis, analysis of the Bragg
peaks provides only information about the average positions, thermal
disorder, and site occupancies. That it, conventional analysis of
Bragg peaks provides a mean field description of the solid. In contrast,
the diffuse part of the scattering contains information about two-point
correlations and therefore can be a rich source of information about
how pairs of atoms or molecules interact with each other. However,
as we briefly mentioned above, extracting information from the diffuse
scattering is still far from routine. Traditionally, two main approaches
have been used to understand the diffuse part of the diffraction pattern
in the kinematic limit: The first is based on a real space formulation
in terms of a set of short-range chemical and displacement pair correlations,
which is specially useful when a small set of correlation parameters
can describe the diffraction pattern and another one based on a reciprocal
space formulation in terms of the modulations of the real space structure
associated with sharp features in \( k \)-space. Obviously, this
later approach is especially useful when sharp features are present
in the diffuse part of the diffraction pattern. Aside from the fact
that some of the approximations involved in the previous approaches
are sometimes difficult to justify (i.e. retaining only the first
few terms of a Taylor expansion), the main problem is that of uniqueness.
As described above, diffuse scattering contains information only about
the pair correlations. Therefore, many different atomic configurations
can lead to the same set of pair correlations (while having very different
higher order correlations, i.e., while representing completely different
configurations) compatible with the set of parameters obtained from
the analysis of the diffraction pattern by using these approaches,
making the determination of the real structure of the material ambiguous.
However, as very well put by Welberry and Butler\cite{Welberry1994},
the ultimate goal of diffraction analysis is to obtain a realistic
model of the structure of the material that is consistent with the
observed diffraction pattern. In this sense, important progress has
been done during recent years by using a method that, in some sense,
is the converse of the aforementioned mathematical modeling. The idea
is to develop a structural model that incorporates all the physical
or chemical constraints already known from the system under consideration.
This model is implemented in the computer (by using Monte Carlo methods)
in such a way that the calculated diffraction pattern qualitatively
agree with the experimental one and quantitative refinements of the
model are iteratively implemented by taking into account the numerical
constraints obtained from the aforementioned mathematical models.
In this way, many unphysical configurations compatible with the diffraction
data are discarded from the outset. In spite of the formidable computational
requirements demanded by this approach, it has already shown its usefulness
in analyzing and obtaining structural information of various systems,
e.g., Fe\( _{3} \)(CO)\( _{12} \)\cite{Welberry1998}, cubic stabilized
zirconia, and disordered organic molecular crystals\cite{Welberry2001,Welberry1994,Welberry1995}.
From these arguments, we can easily understand the importance of complementing
conventional diffraction data with local probes, such as X-ray absorption
fine structure (XAFS) measurements\cite{Conradson1998} and pair distribution
functions (PDF). The long-range average order can be determined by
using conventional analysis of the Bragg peaks while the local structure
can be determined by using XAFS and PDF. This information about the
local structure around each element can be used for constructing physical
structural models that can be provided as input for the Monte Carlo
methods for analyzing the mesoscale structural information contained
in the diffuse scattering and to obtain pair distribution functions.

The necessity of using the previous hierarchical approach to the problem
of determining the real structure of solids has become even more evident
with the increasing realization that local lattice distortions may
act collectively to produce nanoscale phase separation and heterogeneity
(multiple different ordered domains), which are essential ingredients
in understanding the physical properties of complex systems\cite{Conradson2002}.
This is the case, for example, of substitutional impurities in solid
solutions\cite{Ice1999}. If the impurity atoms attract or repel each
other, heterogeneous nanoscale domains or texture will be formed in
the solid. Additional examples of these kinds of effects are charge
separation and charge ordering in colossal magnetoresistance compounds
high \( T_{c} \) superconductors leading to the formation of charge
density waves and stripes\cite{Xu2000,Zaanen2000,Moreo1999,Dagotto2001,Tranquada1995,Tranquada2002}.
These effects give rise to ordered electronic distributions that are
easily observed in diffraction experiments as superlattice or satellite
peaks. 

However, the situation is radically different in the presence of aperiodic
phase separation below the limit of {}``long'' range order, i.e.,
when the structure of the solid cannot be simply described by the
average positions of the atoms plus a small number of distortions.
It is then necessary to consider a large number of correlation functions
characterizing distortions and organization in many different length
scales. In this case, the effects of heterogeneity, organization,
and competition between different phases in the diffraction patterns
are simply not known.

It may seem surprising that, in spite of the fact that it is widely
accepted that some of the most compelling properties of real materials
come from departures from perfect periodicity%
\footnote{See for example Ref.~\onlinecite{Ice1999} and references therein.
}, there is a vast amount of literature concerned with describing theoretically
how different types of disorder affect the diffuse part of the scattering%
\footnote{See for example the extensive bibliography presented in Ref.~\onlinecite{Welberry1995}.
}, and that there is an increasing body of evidence\cite{Bianconi1996,Saini2001,Bianconi2000,Villella2001,Espinosa2001,Espinosa2000,Tyson1996,Li1995,Cox1995,Li1994,Dmowski1995,Vasiliu1999,Shimomura1999}
that suggest that the traditional interpretative frameworks based
on the crystal picture of solids are not enough to understand the
properties of complex systems, the signatures of disorder clearly
seen in experimental diffraction data are, oftentimes, ignored.

One possible reason for this is that, sometimes, the situation in
solids where nanoscale heterogeneity plays an important role can be
even more complicated and misleading than the previous description
may suggest and it is easier to rely on a well-established description
of solids rather than carry on the difficult and expensive program
presented above. However, as we will prove in the present work, ignoring
apparently small and unimportant features usually present in the diffuse
scattering can lead us to develop models and extract conclusions from
these models for systems that, simply put, have nothing to do with
the real solid under consideration. Moreover, we will also show that,
even in some extreme cases, aperiodic distortions affecting one fifth
or more of the atoms in the lattice can barely affect the calculated
structure factors when compared with their periodic counterparts,
making it very easily to miss them by using conventional analysis
methods. We will also see that even in some cases where distinct features
of disorder are present, we can easily devise different types of distorted
configurations that lead to qualitatively similar structure factors,
making it very difficult to ascertain the nature of the underlying
lattice. In order to develop this program, we will present the results
of simulations of the structure factors and radial distribution functions
(RDF) in lattices with different kinds of isolated and collective
distortions. For simplicity, we have carried on the calculations in
two-dimensional lattices. However, we do expect the conclusions drawn
in this work to be extensible to the real three-dimensional case and
the results of such an analysis will be reported elsewhere.

The remainder of this paper is organized as follows: In the next Section
we will describe the lattices we have considered in this work. In
Section III the results of our calculations for the RDF and structure
factors, and a discussion of the main results are presented. Finally,
Section IV is devoted to stating the conclusions of this work.

\section{The distorted lattices}

\begin{figure*}[ht!]
{\centering \subfigure[\label{fig_displacements_a}Randomly selected atoms]{\includegraphics{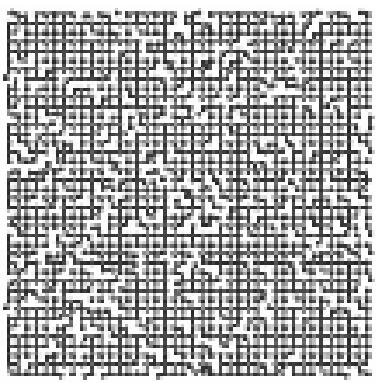}} \hfill{}\subfigure[\label{fig_displacements_b}Channel of displacements]{\includegraphics{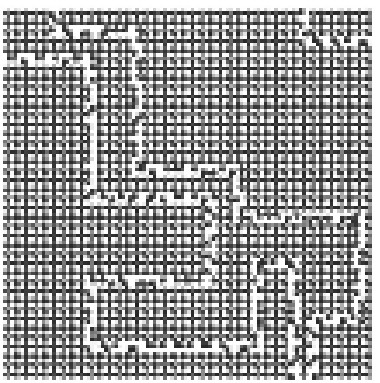}} \hfill{}\subfigure[\label{fig_displacements_c}Atoms forming regular shapes which are shifted as described in the text]{\includegraphics{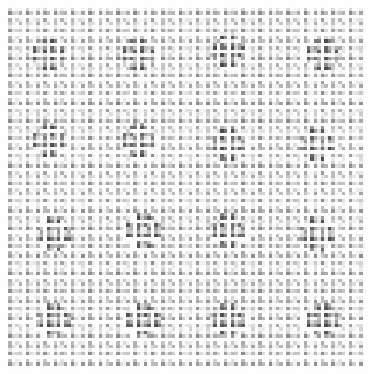}} \hfill{}\subfigure[\label{fig_displacements_d}Atoms forming regular shapes randomly placed in the lattice which are shifted as described in the text]{\includegraphics{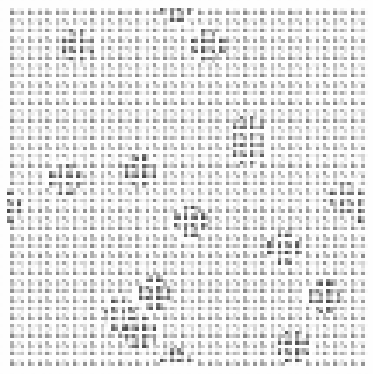}} \par}

\caption{\label{fig_displaced}Lattices with atoms displaced from their crystallographic
positions. As explained in the text, we have also considered two additional
cases not depicted here, analogous to cases (c) and (d), where the
small crosses are additionally contracted an amount \protect\( 0.04a\protect \)
in the horizontal direction.}
\end{figure*}

\begin{figure*}[ht!]
{\centering \subfigure[Single tetragonal domains]{\includegraphics{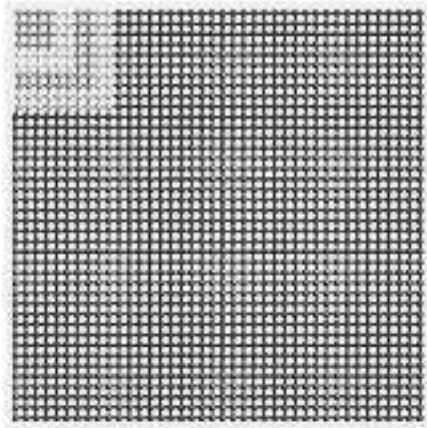}} ~~~~~~~~~~\subfigure[\label{fig_tetradomain_b}Randomly placed tetragonal domains]{\includegraphics{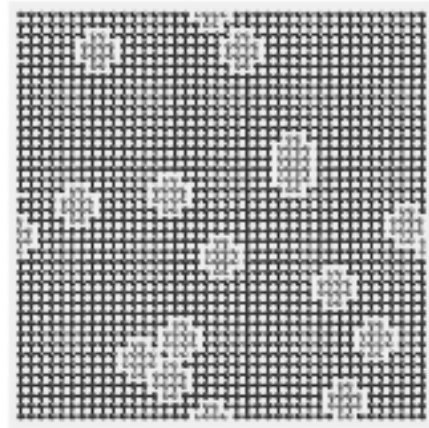}} \par}

\caption{\label{fig_tetradomain}Inclusion of tetragonally distorted domains
in the square lattice. In (a), we have shaded with different grays
some of the domains considered in this work, in particular, \protect\( 4\times 4\protect \),
\protect\( 6\times 6\protect \), \protect\( 8\times 8\protect \),
and \protect\( 10\times 10\protect \) atom domains. In (b), the interatomic
distances of the atoms belonging to the small crosses are contracted
by an amount \protect\( 0.04a\protect \).}
\end{figure*}

\begin{figure*}[ht!]
{\centering \subfigure[\label{fig_modulated_a}Periodic modulated distortions.]{\includegraphics{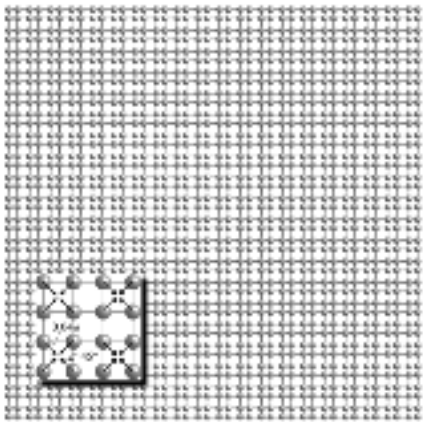}} ~~~~~~~~~~\subfigure[\label{fig_modulated_b}Aperiodic modulated distortions.]{\includegraphics{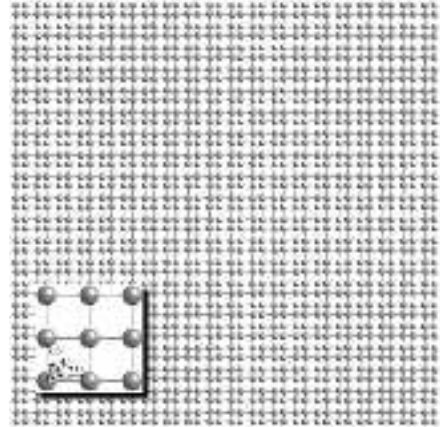}} \par}

\caption{\label{fig_modulated}Modulated lattices. The insets show how the
distortions are implemented in both the periodic and aperiodic cases.}
\end{figure*}

\begin{figure*}[ht!]
{\centering \subfigure[\label{fig_interstitial_a}Periodically distributed interstitial atoms]{\includegraphics{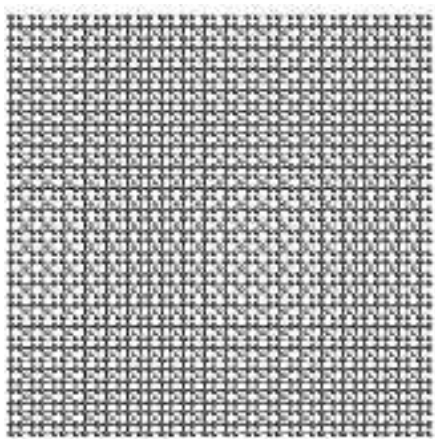}} ~~~~~~~~~~\subfigure[\label{fig_interstitial_b}Aperiodically distributed interstitial atoms]{\includegraphics{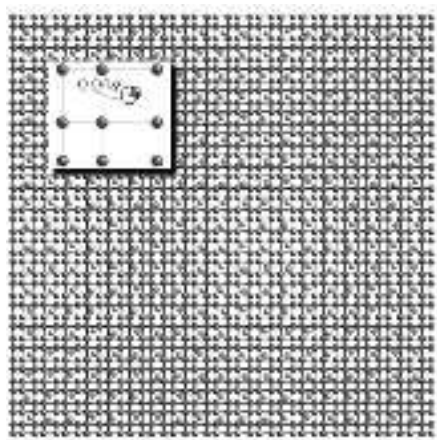}} \par}

\caption{\label{fig_interstitial}Lattices with interstitial atoms. The inset
in figure (b) shows how the interstitial atoms are placed on the lattice
in the aperiodic case. They are placed on a randomly chosen point
of the circumference with center at the center of the cell and radius
\protect\( 0.04a\protect \).}
\end{figure*}

\begin{figure*}[ht!]
{\centering \subfigure[\label{fig_channels_a}Ordered periodic channels]{\includegraphics{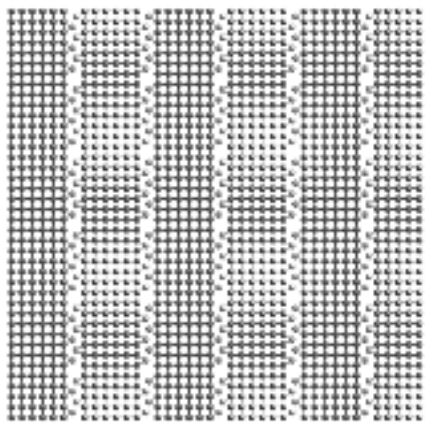}} ~~~~~~~~~~\subfigure[\label{fig_channels_b}Aperiodic channels]{\includegraphics{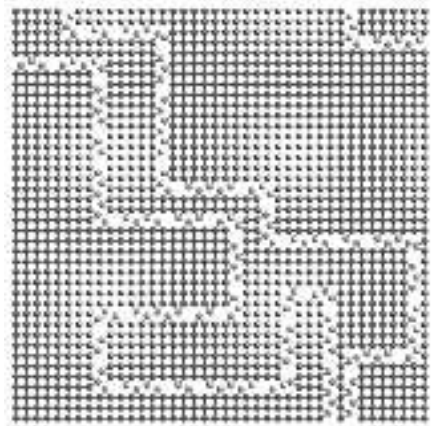}} \par}

\caption{\label{fig_channels}Lattices with channels of tetragonal distortions.}
\end{figure*}

As briefly mentioned in the Introduction, we have considered 2D square
lattices to which different kinds of static disorder have been applied
in order to simulate various possible lattice distortions. For completeness,
we have also considered purely periodic lattices obtained by distorting
the square lattice (see below). The size of all the lattices reported
in this work is \( 40\times 40 \) atoms (1600 atoms). This is a compromise
between large enough lattices so the finite size effects are small
while, at the same time, keeping the CPU time down to a reasonable
amount (hours). Obviously, the finite size of the lattices will give
raise to a broadening of the peaks in the calculated structure factors
that makes it difficult to extract any conclusion about how these
distortions affect the shape of the diffraction peaks. However, such
an analysis of the shape of the peaks is outside the scope of this
work and, besides, based on comparisons with the results for smaller
lattices, this broadening does not modify any of the conclusions reached
in this work about the effects of distortions on the diffuse scattering.
The lattice parameter used in all the calculations is 3 Å. Let us
now briefly review the different kinds of distortions that have been
applied to the otherwise perfect square lattice:

\begin{itemize}
\item The first case we have considered consists of different configurations
of vacancies placed on the square lattice (see Fig. \ref{fig_vacancies}).
The first and simplest configuration of vacancies consists on randomly
removing 20\% of the atoms of the lattice in order to test the effect
of random, aperiodic elimination. In addition, we have also considered
configurations in which the positions of the vacancies are not random,
but correlated, forming channels across the lattice and periodic and
aperiodic distributions of vacancies defining certain regular geometric
shapes. Finally, we have also considered a configuration in which
the vacancies define different regular shapes which are randomly placed
in the lattice.
\item Atoms displaced from their original crystallographic positions (Fig.~\ref{fig_displaced}).
The configurations considered in this case are similar to the previous
ones. The first type of displacements we have considered consists
on randomly choosing a 20\% of the atoms of the lattice and displacing
them by a vector \( \left( \cos \theta ,\sin \theta \right)  \),
where \( \theta  \) randomly varies among \( 32^{\circ } \), \( 148^{\circ } \),
\( 212^{\circ } \), and \( 328^{\circ } \) (see Fig.~\ref{fig_displacements_a}).
The second type of displacements we have considered is similar to
the channel of vacancies explained above, except for the fact that
now, instead of removing the atoms, they are displaced by the same
random vector as in the previous displaced configuration (see Fig.~\ref{fig_displacements_b}).
We have also considered the case in which a certain number of atoms
defining a regular shape (a small cross) are displaced by the same
random vector, and this shape is periodically and aperiodically repeated
across the lattice (see Figs.~\ref{fig_displacements_c} and \ref{fig_displacements_d}).
Finally, we have also tested the effect of having two phases of different
symmetries inside the lattice. To this end, the interatomic distances
between the atoms defining the crosses used in the last two configurations
have been contracted by a factor \( 0.04a \) in the horizontal direction.
These two configurations have not been depicted, at the visual differences
with respect to the last two cases are barely noticeable. 
\item Inclusion of a tetragonal domain. We have considered the effect of
contracting a regular domain (a square domain) in the horizontal direction
by an amount \( 0.04a \) (see Fig.~\ref{fig_tetradomain}). In order
to see what is the minimum size this domain has to be, in order to
show up in the \( S(Q) \), different domain sizes have been considered,
ranging from \( 4\times 4 \) atoms up to \( 40\times 20 \) atoms
(half the lattice is contracted). Additionally, we have also considered
a lattice in which the positions of groups of atoms defining a geometric
motif (a small cross), randomly placed in the lattice, have been tetragonally
distorted (see Fig.~\ref{fig_tetradomain_b}).
\item Periodic and aperiodic modulated distortions (Fig.~\ref{fig_modulated}).
Each of the elementary square cells in the lattice is assigned a position
vector of the form \( (i,j) \), where \( i \) and \( j \) are both
integer numbers. For example, the (1,0) cell would correspond to the
cell defined by the atoms located at (0,0), (0,\( a \)), (\( a \),0),
and (\( a \),\( a \)). In order to implement this kind of distortion,
in the periodic case, the positions of all the atoms on the corners
of the cells with odd \( i \) and \( j \) have been shifted towards
the center of the corresponding cell by a 4\% of the undistorted lattice
parameter (see Fig.~\ref{fig_modulated_a} for details). The resulting
lattice can be seen as another square lattice of lattice parameter
\( 2a \) with a basis. In the aperiodic case (Fig.~\ref{fig_modulated_b}),
the procedure is similar to the periodic case, but we have allowed
the direction of the contraction to randomly vary among the values
\( 25^{\circ } \), \( 45^{\circ } \), and \( 65^{\circ } \).
\item Periodic and aperiodic interstitial atoms. Using the periodic modulated
lattice described in the previous paragraph, an interstitial atom
has been placed inside each of the squares defined by those atoms
with largest bondlength (see Fig.~\ref{fig_interstitial}). In the
periodic case, the interstitial atom is located in the center of the
square. Again, this new lattice can be seen as a square lattice of
lattice parameter \( 2a \) with a basis. In the aperiodic case, the
interstitial atom is randomly located in a circle of radius \( 0.04a \)
centered at the middle point of the square (see Fig.~\ref{fig_interstitial_b}
for details).
\item Periodic and aperiodic channels of defects across the lattice. In
this case, the effect of correlated networks of defects has been studied.
In the periodic case, all the atoms belonging to a chosen column have
been shifted downwards by half the lattice parameter. At the same
time, the atoms belonging to this channel have been shifted in the
horizontal direction by an amount \( \frac{a}{6} \) or \( -\frac{a}{6} \),
respectively (see Fig.~\ref{fig_channels_a}). In addition, each seventh
atom in the network has been removed. Finally, the distorted network
has been periodically repeated across the horizontal direction. In
the aperiodic case, channels across the lattice similar to the ones
introduced in the case of vacancies and displacements have been considered
(see Fig.~\ref{fig_channels_b}). However, in contrast with the displacements
case, the atoms inside this channel have been alternatively displaced
horizontally or vertically, in such a way that they define a second
tetragonal phase inside the square lattice, whereas in the displacements
case, the displacements where applied in a random fashion. Moreover,
each seventh atom has been removed.
\end{itemize}
As stated in the Introduction, our intention is to study how these
different types of distortions affect the calculated diffraction patterns.
The result of this analysis is presented in the next Section.

\section{Results and discussion of the calculated structure factors and radial
distribution functions}

The static structure factor (SF) \( S(Q) \) and radial distribution
function (RDF) \( g(r) \) have been calculated for the lattices introduced
in the previous Section. To this end, we have used the code CRYGEN2D,
developed by P. Villella and coworkers at Los Alamos National Lab
\cite{Villela2002}. This program takes as input the coordinates of
the atoms in a two-dimensional lattice and exactly computes the RDF
and \( S(Q) \). For comparison, the \( S(Q) \) and \( g(r) \) of
a \( 40\times 40 \) atoms square lattice have been also calculated,
and used as reference. In order to simulate the effects of finite
temperature, we have used a typical Debye-Waller vibrational factor
of amplitude 0.05 Å. The \( S(Q) \) data have been normalized in
order to eliminate differences coming from the different number of
atoms present in some lattices with respect to the square one. This
allows us to compute the difference between any two \( S(Q) \) data
sets, so the differences in the diffuse scattering are more noticeable.
The results of the \( S(Q) \) calculations are presented below in
terms of the dimensionless variable \( \left( \frac{Q\, a}{2\pi }\right) ^{2}=h^{2}+k^{2} \)
(with \( h \) and \( k \) the Miller indices), which makes it very
simple to assign the Bragg peaks. Let us now review how these different
kinds of distortions affect the calculated \( g(r) \) and \( S(Q) \).

\noindent \emph{Effect of random vacancies.-} The results for the
RDF and \( S(Q) \) can be seen in Fig. \ref{fig_vacancies_soq}.
As one would expect, the only effect of vacancies on the RDF is a
reduction in the intensity of the peaks proportional to the concentration
of vacancies in the lattice.

The situation is a little more complicated for the calculated structure
factors. The fundamental peaks are located at the same positions as
for the square lattice and have the same shapes. Their intensities,
however, are smaller for the lattices with vacancies than for the
square lattice due to the fact that part of the intensity goes now
to the diffuse part of the scattering. For the case of 20\% atoms
randomly removed from the lattice, the main effect is an increment
of the background, very similar to the effect of a Debye-Waller factor,
except for the fact that this background is essentially constant for
all \( Q \)'s, in contrast to the Debye-Waller diffuse scattering,
which is larger for larger values of \( Q \). In the case of the
channels of vacancies extending through the lattice, the main effect
is an increment of the diffuse scattering around the \( [i\, 0]\, \, (i=1,2,3,\ldots ) \)
peaks. This enhancement is highly asymmetric, affecting only the high-\( Q \)
part of the peak.

\begin{figure*}[ht!]
{\centering \subfigure[\label{fig_vacancies_a}Vacancies randomly placed in the lattice]{\includegraphics{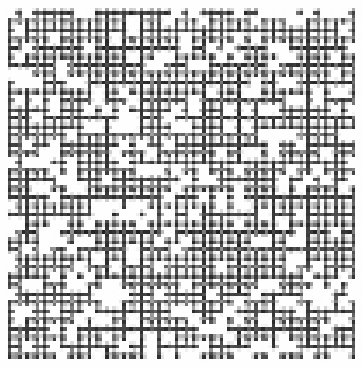}} \hfill{}\subfigure[\label{fig_vacancies_b}Vacancies forming channels accross the lattice]{\includegraphics{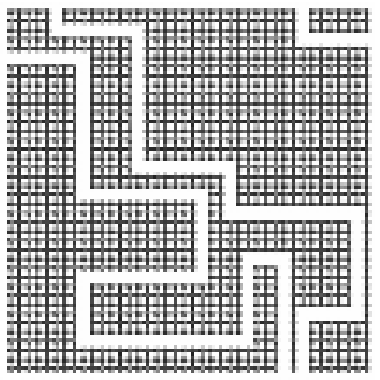}} \hfill{}\subfigure[\label{fig_vacancies_c}Vacancies forming a regular shape which is distributed in an ordered fashion]{\includegraphics{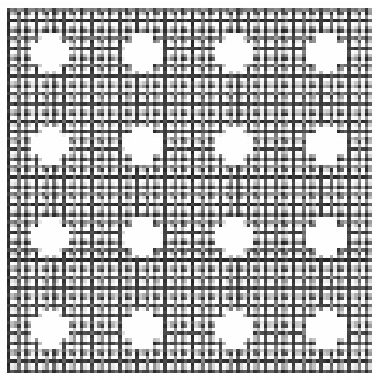}} \par}

{\centering \hfill{}\subfigure[\label{fig_vacancies_d}Vacancies forming a regular shape which is randomly placed on the lattice]{\includegraphics{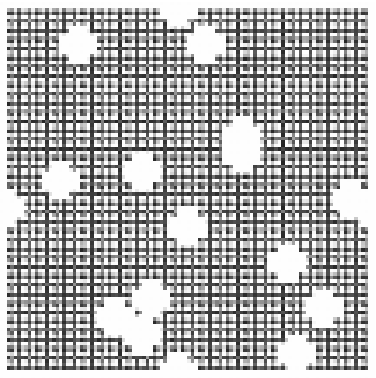}} \hfill{}\subfigure[\label{fig_vacancies_e}Vacancies forming different shapes which are randomly placed on the lattice]{\includegraphics{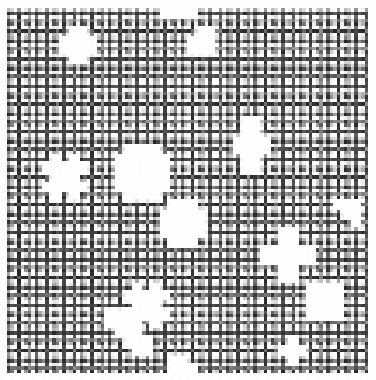}} \hfill{} \par}

\caption{\label{fig_vacancies}The configurations of vacancies considered
in this work.}
\end{figure*}

\begin{figure*}[ht!]
{\centering \subfigure[\label{fig_vacancies_soq_a}Radial distribution function]{\resizebox*{0.39\textwidth}{!}{\includegraphics{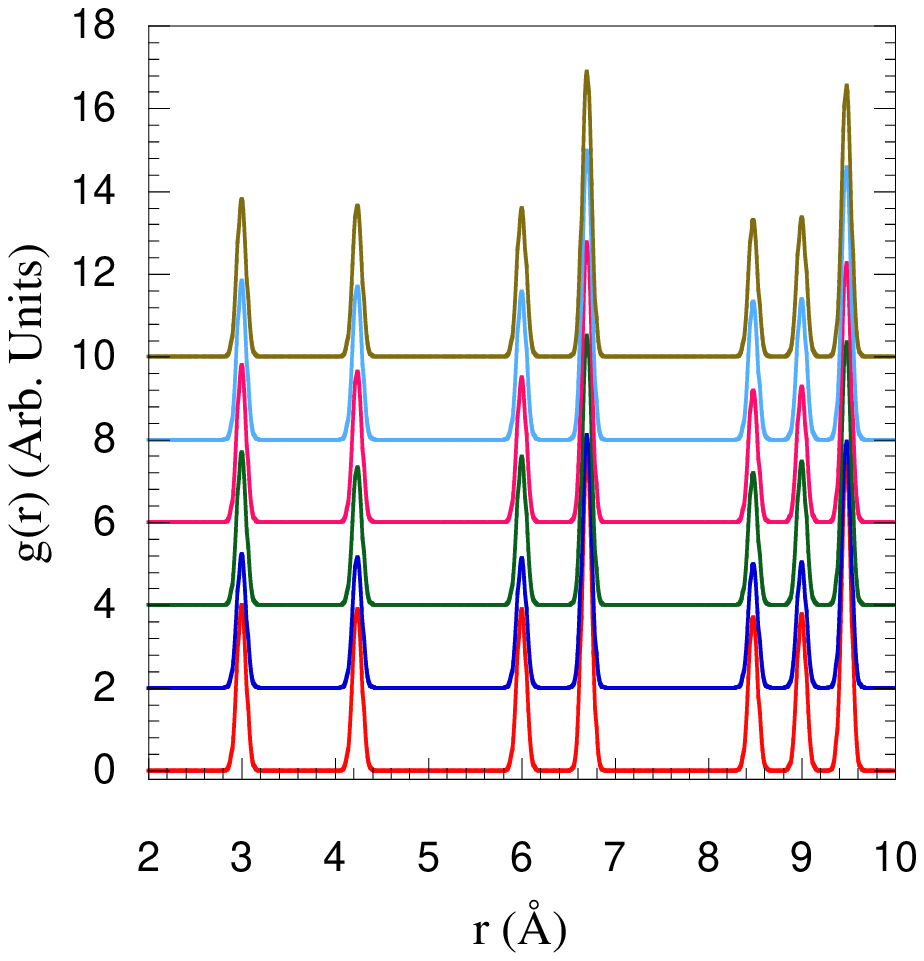}}} \hfill{}\subfigure[\label{fig_vacancies_soq_b}Structure function]{\includegraphics{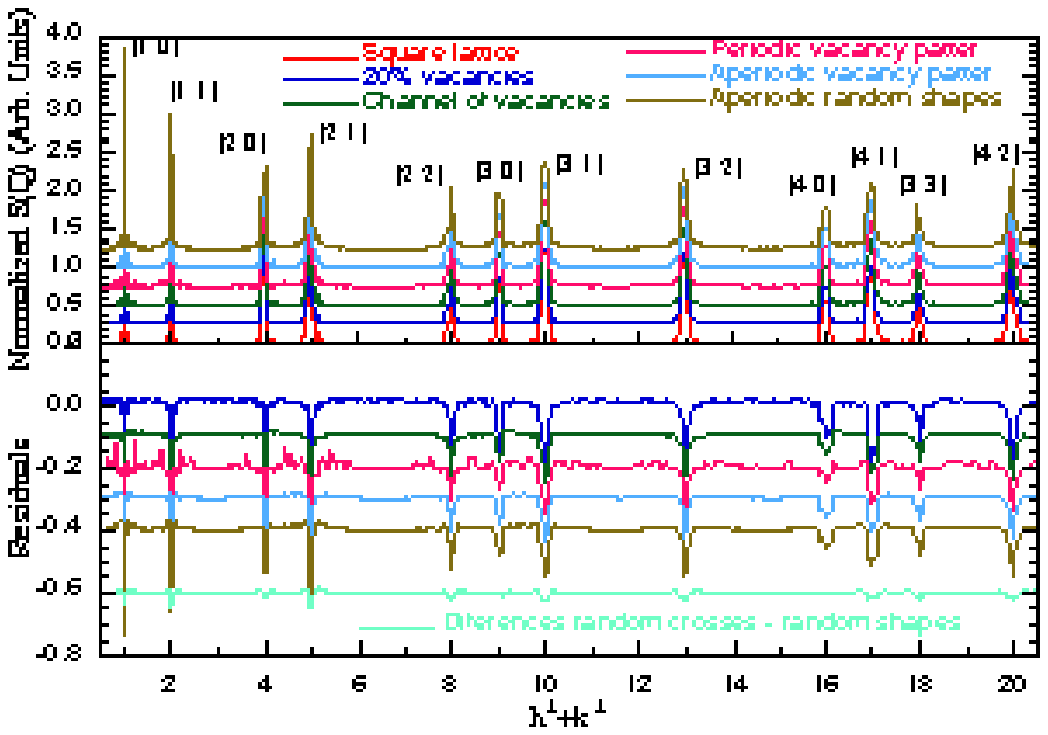}} \par}

\caption{\label{fig_vacancies_soq}RDF and \protect\( S(Q)\protect \) for
the lattices with vacancies. The assignments of the diffraction peaks
of the square lattice are shown in the upper panel of (b) for reference.
The lower panel shows in greater detail the difference between the
\protect\( S(Q)\protect \) of the square lattice and the ones of
the distorted lattices. In order to construct these curves, the \protect\( S(Q)\protect \)
have been properly normalized and the \protect\( S(Q)\protect \)
of the square lattice has been subtracted from the rest of \protect\( S(Q)\protect \)
of the distorted lattices. The units used in that panel are the same
as the ones used in the upper panel. In this way, negative values
correspond to regions in which the \protect\( S(Q)\protect \) of
the square lattice is larger than the corresponding one for the distorted
lattice. As expected, peaks with negative values occur at the positions
of the fundamental peaks of the square lattice diffraction pattern,
whereas peaks with positive values correspond to additional features
coming from the distorted character of the lattices (superlattice
peaks or diffuse scattering). The green curve labeled {}``Differences
random\ldots{}'' has been obtained by subtracting the \protect\( S(Q)\protect \)
of the lattice in Fig.~\ref{fig_vacancies_e} from the one of the
lattice depicted in Fig.~\ref{fig_vacancies_d}.}
\end{figure*}

If we now turn our attention to the configurations in which vacancies
define geometric shapes, we can see that, in the periodic case, many
new peaks appear at incommensurate positions with respect to the spacing
of the original square lattice. Obviously, these new peaks are nothing
but superlattice peaks associated with the fact that the lattice can
still be seen as a square lattice of lattice parameter \( 10a \)
with a complicated basis. We will further comment on these superlattice
peaks below. In contrast, if these shapes are randomly distributed,
the superlattice peaks disappear, and the background acquires an modulated
character: there is an enhanced diffuse scattering around the fundamental
peaks, and this background is depleted in between. The combination
of these two effects give rise to some kind of oscillatory diffuse
scattering. This effect is more noticeable at low \( Q \). We wanted
to check if these oscillations were due to the fact that all the defects
have the same size and shape and, to this end, we repeat this calculation
using different shapes and sizes, but located at the same positions
(see Fig.~\ref{fig_vacancies_e}). As can be seen from the green curve
in the lower panel of Fig.~\ref{fig_vacancies_soq_b}, this is not
the case, and the only difference between these two configurations
is a barely noticeable increment of the diffuse scattering around
the fundamental peaks at larger \( Q \), in this last case.

\begin{figure*}[ht!]
{\centering \subfigure[\label{fig_displacements_soq_a}Radial distribution function]{\resizebox*{0.39\textwidth}{!}{\includegraphics{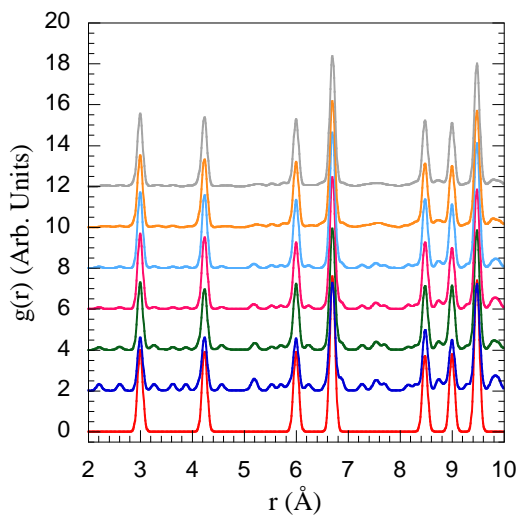}}} \hfill{}\subfigure[\label{fig_displacements_soq_b}Structure factor]{\includegraphics{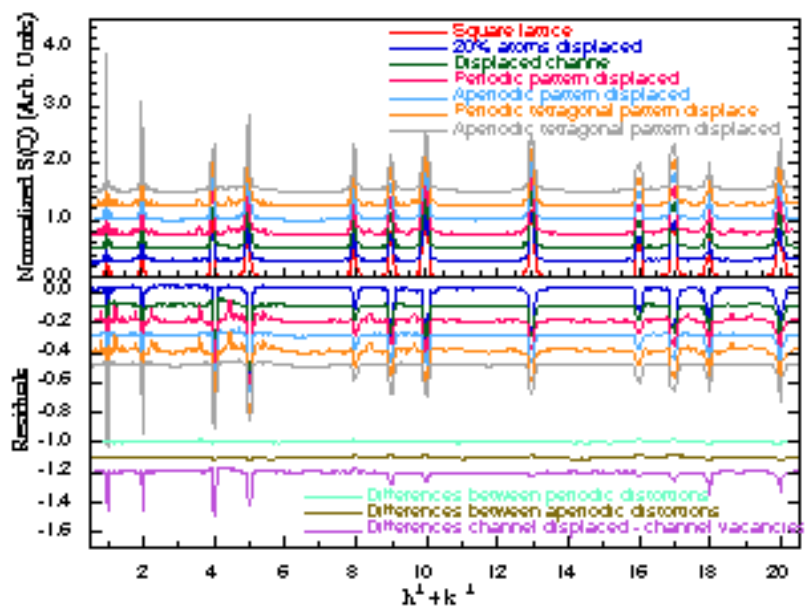}} \par}

\caption{\label{fig_displacements_soq}RDF and \protect\( S(Q)\protect \)
for the lattices with displacement distortions. See the comments in
the caption of figure \ref{fig_vacancies_soq} about how the lower
panel has been constructed. The curve labeled {}``Differences between
periodic distortions'' has been calculated by subtracting the \protect\( S(Q)\protect \)
of the lattice depicted in Fig.~\ref{fig_displacements_c} to the
one of the analogous lattices where additional tetragonal contractions
have been applied to the small crosses distributed in an ordered fashion
in the lattice. The same procedure has been applied to construct the
brown curve in that panel but using the analogous lattices where the
small geometrical motifs are randomly placed.}
\end{figure*}

\noindent \emph{Effect of aperiodic displacements.-} The results for
this case can be seen in Fig.~\ref{fig_displacements_soq}. The RDF
for the distorted lattices shows the original peaks of the square
lattice (the intensity of these peaks is reduced, as expected) plus
additional new peaks due to the new pairwise distances introduced
by the displaced atoms. The new distances are especially apparent
in the RDF for the lattice in which 20\% of the atoms are randomly
chosen and shifted from their crystallographic positions. For example,
if we consider the peak corresponding to the nearest neighbor (NN)
distances, we can see that four new peaks corresponding to the four
new NN distances introduced by the distortions. The intensity of these
new peaks are the same, as each of the four displacement vectors are
equally probable. Similar comments apply to further neighbor peaks.
The other configurations exhibit exactly the same features as this
last one: The same new peaks appear and they are located at the same
positions. The main difference is the intensity of the new peaks which,
obviously, is higher for those configurations involving a larger number
of atoms displaced (320 atoms in the case of the 20\% randomly chosen
atoms to be compared with 200 in the channel case and 192 for the
shifted crosses, respectively). This reduction in the intensity of
the peaks is especially noticeable in those configurations in which
the displaced atoms define a regular shape, as in this case the only
new distances are located at the interface between the displaced and
undistorted lattice. When additional tetragonal contractions are applied
to these shapes, there are also new distances between the atoms defining
the shape itself (for example, the NN distance in the horizontal direction
is now 2.88 Å). However, these new distances are masked by the thermal
broadening of the peaks associated with the regular part of the lattice.

The situation is more complicated for the calculated \( S(Q) \).
The intensity of the fundamental peaks decreases as part of it goes
into additional structure appearing in the diffuse part of the scattering.
For the configuration in which 20\% of the atoms are randomly chosen
and shifted, the background scattering exhibits modulated variations
(enhanced background at low \( Q \), depleted at intermediate values,
and enhanced again at high values of \( Q \)), giving the impression
of an oscillatory background, clearly different from the effect of
purely thermal disorder. For the channel of displaced atoms, the behavior
of the diffuse scattering is very similar to that of the channel of
vacancies discussed above: The high-\( Q \) part of the \( [i\, 0]\, \, (i=1,2,3,\ldots ) \)
reflections is enhanced. For comparison, we have depicted the difference
between the \( S(Q) \) for this configuration and the one for the
channel of vacancies (the violet curve in the bottom panel of Fig.~\ref{fig_displacements_soq_b}).
As can be seen from that curve, the only difference between the \( S(Q) \)
for these two configurations is the intensity of the fundamental reflections.
For other values of \( Q \) this curve is essentially a flat line.

If we now turn our attention to the configurations where the shifted
atoms define geometrical shapes, we can observe from Fig~\ref{fig_displacements_soq_b}
that the results for those cases in which the regular shapes are periodically
chosen and then randomly shifted (both with and without tetragonal
contraction) are very interesting: The resulting \( S(Q) \) resembles
that of a periodic lattice (appearance of superlattice peaks at incommensurate
\( [h\, k] \) values), even though there does not exists an elementary
motif (unit cell) from which the whole lattice can be obtained by
periodic repetition. When these shapes are randomly distributed, the
resulting diffuse scattering lacks any sharp features and, instead,
it exhibits an oscillatory character with enhanced diffuse background
around the fundamental reflections and depleted in between. Strikingly,
the calculated SFs for both the regular shapes with and without tetragonal
contractions are essentially identical, as can be seen from the green
and brown curves in the bottom panel of Fig.~\ref{fig_displacements_soq_b},
which are essentially flat lines. These observations imply that, even
though we can induce the presence of a random or ordered motif of
defects, it is difficult to judge whether this motif contains no atoms
at all or a phase with a symmetry completely different from the one
of the initial lattice. We will further develop this point when stating
the conclusions of this work.

\begin{figure*}[ht!]
{\centering \subfigure[\label{fig_tetradomain_soq_a}Radial distribution function]{\resizebox*{0.39\textwidth}{!}{\includegraphics{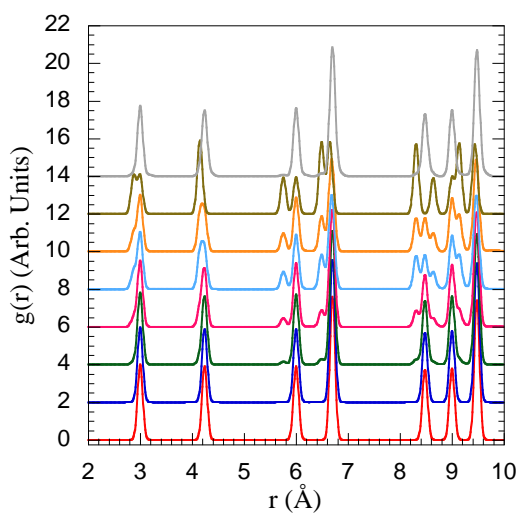}}} \hfill{}\subfigure[\label{fig_tetradomain_soq_b}Structure factor]{\includegraphics{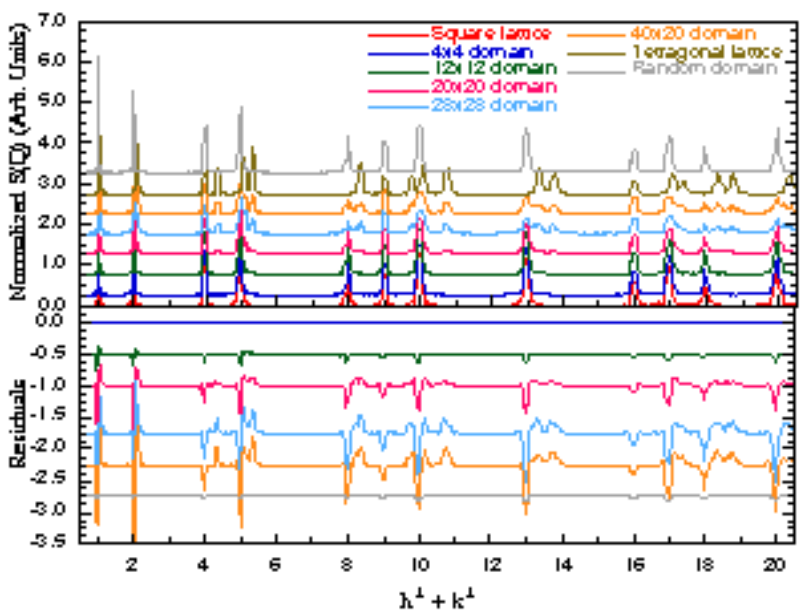}} \par}

\caption{\label{fig_tetradomain_soq}RDF and \protect\( S(Q)\protect \) for
the lattice with a tetragonal domain. See the comments in the caption
of figure \ref{fig_vacancies_soq} about how the lower panel has been
constructed.}
\end{figure*}

\noindent \emph{Effect of the inclusion of a tetragonal domain.-}
The results of the calculations for this type of distortions can be
seen in Fig.~\ref{fig_tetradomain_soq}. In that figure, we have
only depicted the results for some representative domain sizes. For
the sake of comparison, the \( S(Q) \) for a \( 40\times 40 \) lattice
tetragonally contracted in the horizontal direction is also shown.
As stated above, we also show in that figure the results RDF and \( S(Q) \)
for the lattice depicted in Fig.~\ref{fig_tetradomain_b}, where randomly
placed regular shapes have been horizontally contracted. The RDF for
those lattices with a single tetragonal domain nicely interpolates
between the one for the square lattice and the one of a purely tetragonal
lattice, and peaks associated to the new distances inside the tetragonal
domain clearly develop as the domain size increases. It is interesting
to note, though, that for practical purposes, below a domain size
of \( 12\times 12 \) atoms, the new distances are barely noticeable
in the RDF. The RDF for the randomly distributed tetragonally contracted
shapes is very similar to the one consisting of a single tetragonal
domain with the same number of atoms with the only difference of some
small and broad peaks occurring at higher distances associated to
the distances between the boundaries of the geometrical shapes and
the atoms in the square phase.

Regarding the \( S(Q) \), we can see in Fig.~\ref{fig_tetradomain_soq_b}
that new peaks, typical of the tetragonal symmetry, develop as the
domain size is increased. Interestingly, these peaks are only noticeable
for large domain size and, even for the \( 40\times 20 \) domain,
the tetragonal peaks are not fully developed. However, the resulting
\( S(Q) \) for this lattice is fully consistent with the one obtained
by adding up the \( S(Q) \) for a square \( 40\times 20 \) lattice
plus the one for a \( 40\times 20 \) one. Furthermore, the positions
of the fundamental reflections are shifted to higher \( Q \), due
to an overall (though small) contraction of the lattice. Even though
we have not shown it in that figure, we found that for sizes smaller
than \( 12\times 12 \) atoms the new peaks are barely noticeable.
The \( S(Q) \) for the lattice with tetragonal shapes randomly distributed
is essentially identical to the one of the square lattice, as can
be seen in the lower panel of Fig.~\ref{fig_tetradomain_soq_b}.

\begin{figure*}[ht!]
{\centering \subfigure[\label{fig_modulated_soq_a}Radial distribution function]{\resizebox*{0.39\textwidth}{!}{\includegraphics{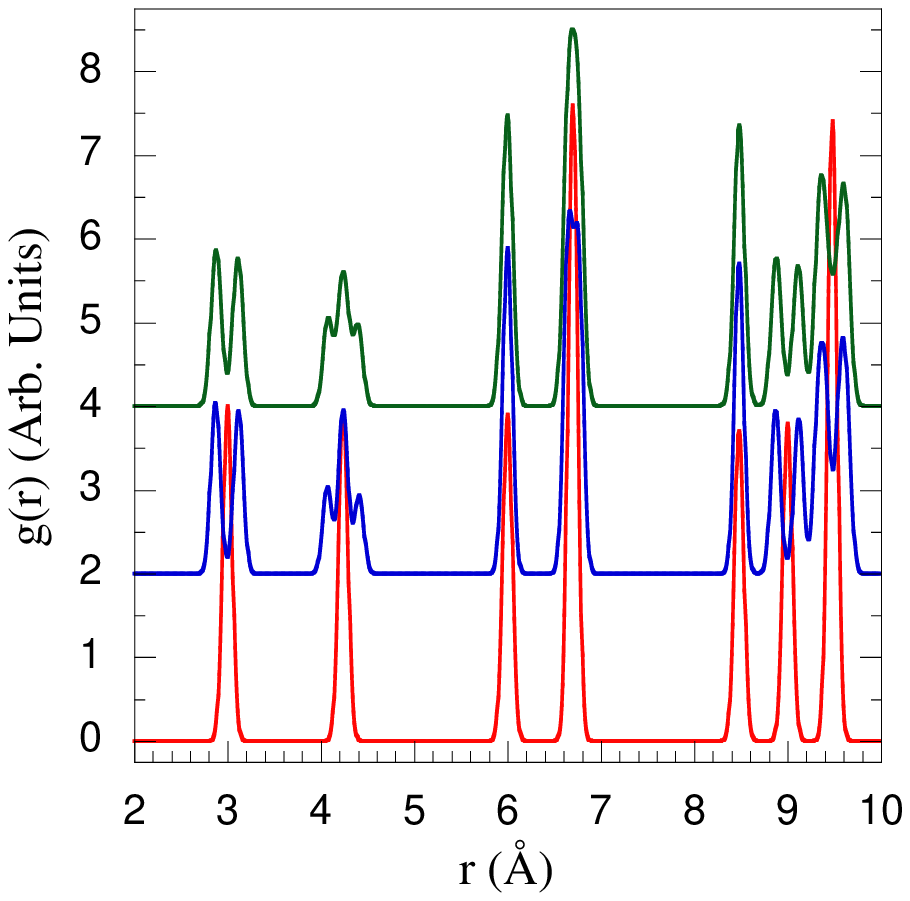}}} \hfill{}\subfigure[\label{fig_modulated_soq_b}Structure factor]{\includegraphics{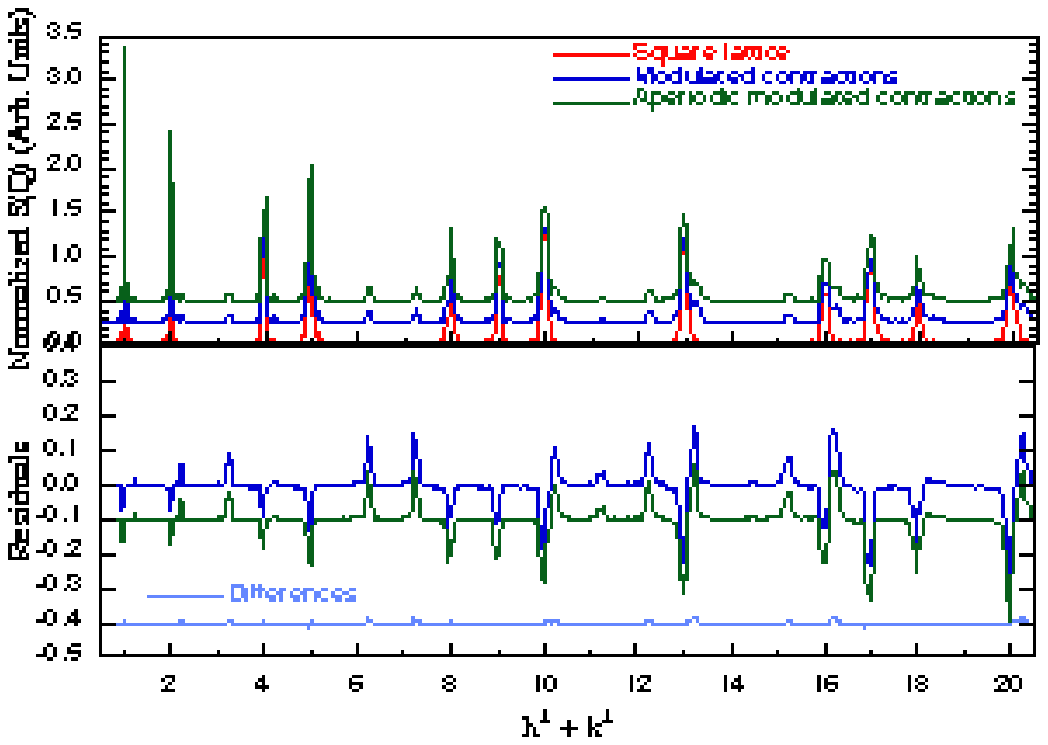}} \par}

\caption{\label{fig_modulated_soq}RDF and \protect\( S(Q)\protect \) for
the modulated lattices. See the comments in the caption of figure
\ref{fig_vacancies_soq} about how the lower panel has been constructed.}
\end{figure*}

\noindent \emph{Effect of periodic and aperiodic modulated distortions.-}
The results of the calculations for this type of distortions can be
seen in Fig.~\ref{fig_modulated_soq}. The RDF for the modulated
lattices (both periodic and aperiodic), are quite distinct from the
square lattice one, as new distances are generated in the distorted
lattices. As a result of these distortions, some of the peaks (noticeably
the ones corresponding to NN, next NN, 6th NN, and 7th NN) are split
reflecting these new distances, whereas in other cases (3rd, 4th,
and 5th NN) the only effect is an additional (small) broadening superimposed
to the thermal one. We can also see that the RDF for the aperiodic
lattice is very similar to the periodic one, as we could expect from
the fact that the distortions we have implemented average out to zero.
The only noticeable difference is an slight additional broadening
of the split peaks.

The \( S(Q) \) for these lattices shows interesting features again.
In the periodic case, many new small superlattice peaks at incommensurate
\( h \) and \( k \) values appear associated with the fact that
the modulated lattice is another square lattice of parameter \( 2a \)
with a basis. The interesting point, however, is that the \( S(Q) \)
for the aperiodic modulated lattice is almost identical to the periodic
one, as can be seen from the curve labeled {}``Differences'' in
the lower panel of Fig.~\ref{fig_modulated_soq_b}. The only noticeable
difference is the appearance of a marginal amount of diffuse scattering
close to the fundamental reflections. Moreover, in both the periodic
and aperiodic cases, the original peaks are only minimally affected
until \( h^{2}+k^{2}\approx 16 \).

\begin{figure*}[ht!]
{\centering \subfigure[\label{fig_interstitial_soq_a}Radial distribution function]{\resizebox*{0.39\textwidth}{!}{\includegraphics{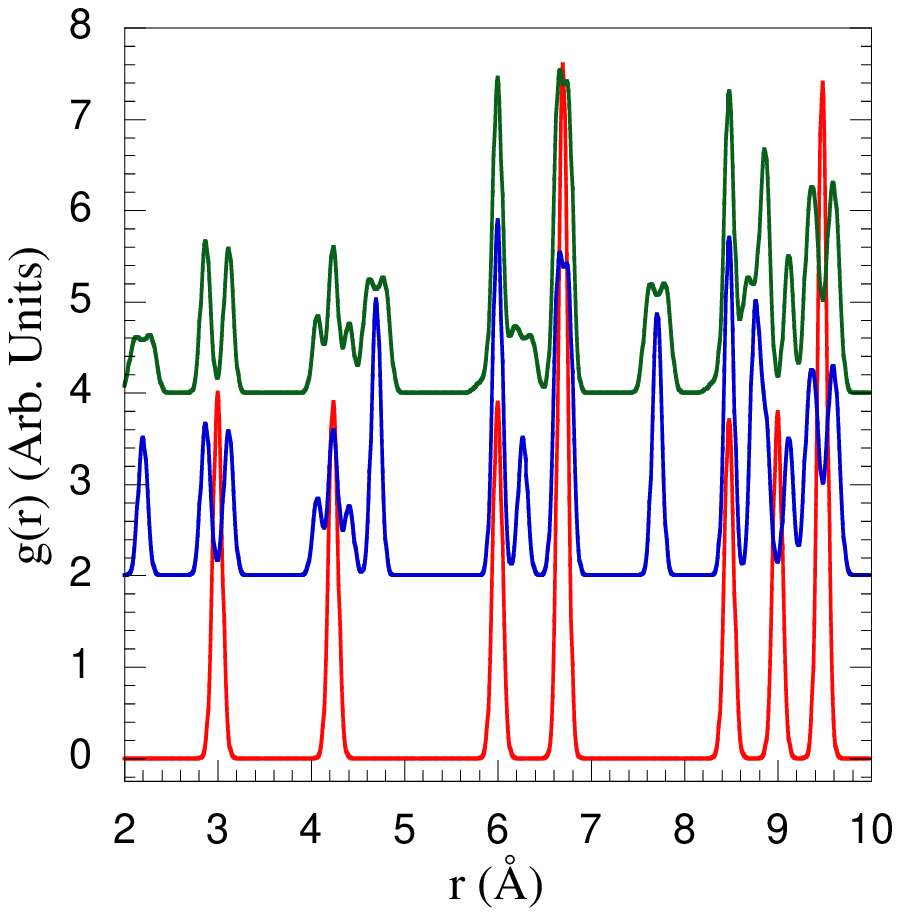}}} \hfill{}\subfigure[\label{fig_interstitial_soq_b}Structure factor]{\includegraphics{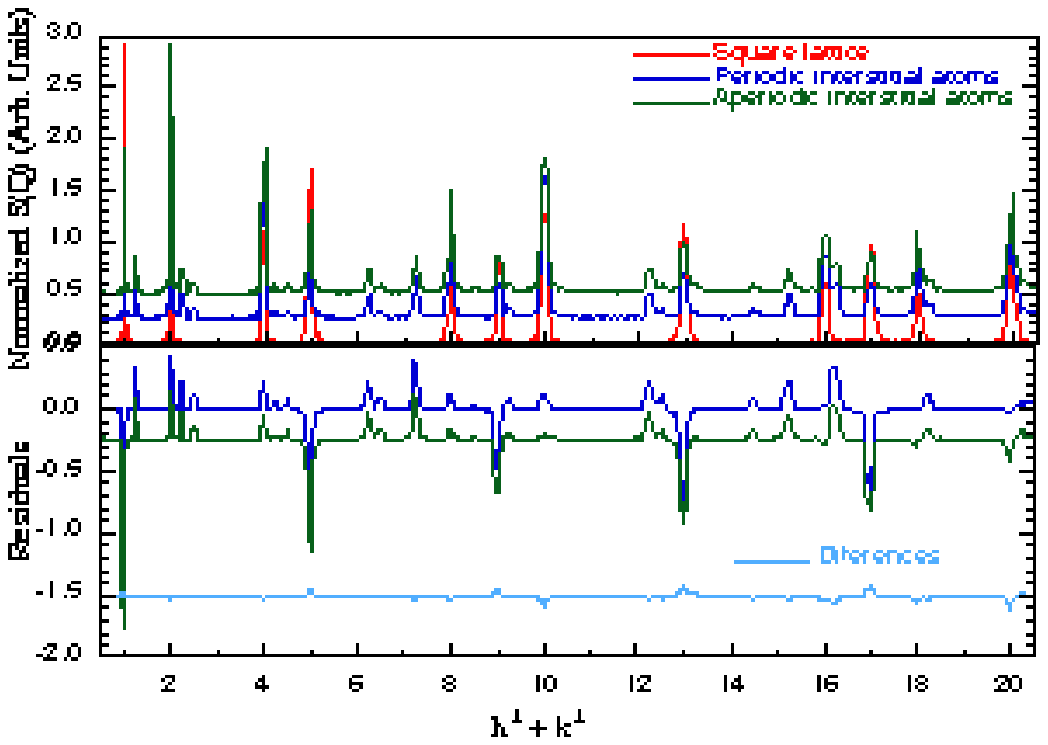}} \par}

\caption{\label{fig_interstitial_soq}RDF and \protect\( S(Q)\protect \)
for the lattices with interstitial atoms. See the comments in the
caption of figure \ref{fig_vacancies_soq} about how the lower panel
has been constructed.}
\end{figure*}

\noindent \emph{Effect of periodic and aperiodic interstitial atoms.-}
The results of the calculations for these types of lattices can be
seen in Fig.~\ref{fig_interstitial_soq}. Regarding the RDF calculation,
new distinct peaks appear associated with the new atom in the basis
(see Fig.~\ref{fig_interstitial_soq_a}), when compared with the calculated
RDF for modulated lattice described above . This is especially clear
in the periodic case. In the aperiodic case, these same peaks remain,
but they are broadened by disorder, as expected.

Regarding the results for the \( S(Q) \), this quantity exhibits
the peaks already described in the modulated lattice case plus new
superlattice peaks associated with the sublattice of interstitial
atoms. However, in full analogy with the previous case, the \( S(Q) \)
for the aperiodic case is essentially identical to the periodic one,
the difference between this two cases being again a slight increment
of the diffuse scattering around the fundamental peaks. When compared
with the same curve for the modulated lattice the only difference
is that the differences in the diffuse scattering around the fundamental
reflections slightly increase at higher \( Q \).

\begin{figure*}[ht!]
{\centering \subfigure[\label{fig_channels_soq_a}Radial distribution function]{\resizebox*{0.39\textwidth}{!}{\includegraphics{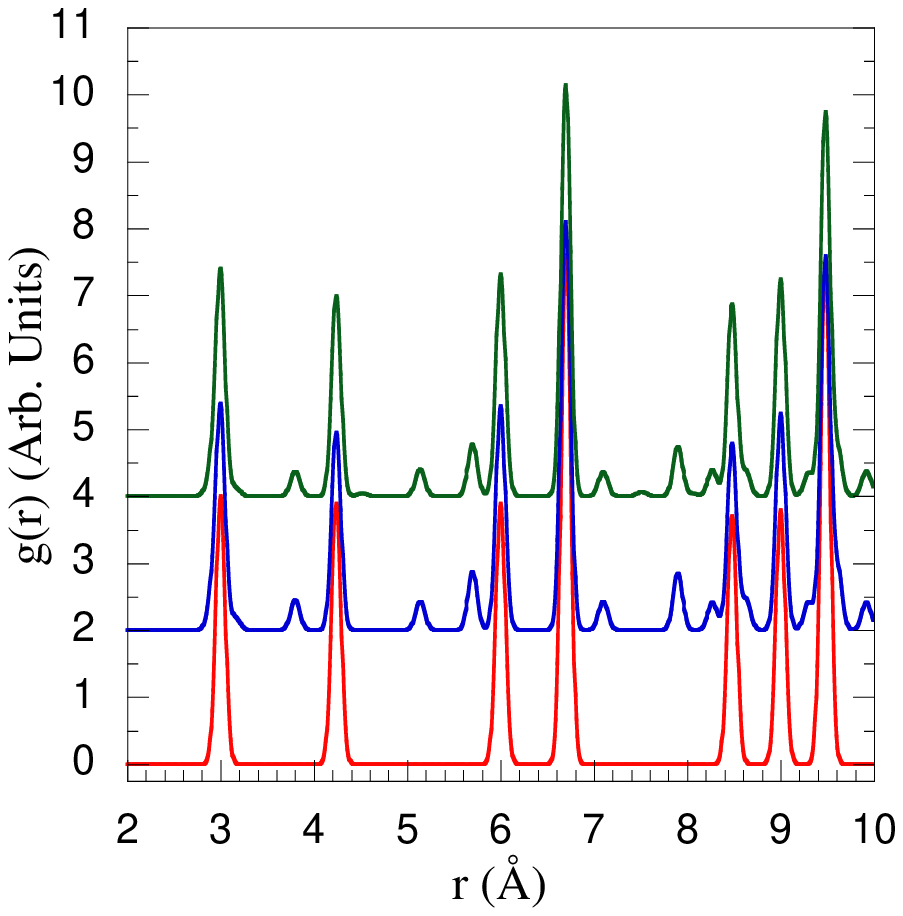}}} \hfill{}\subfigure[\label{fig_channels_soq_b}Structure factor]{\includegraphics{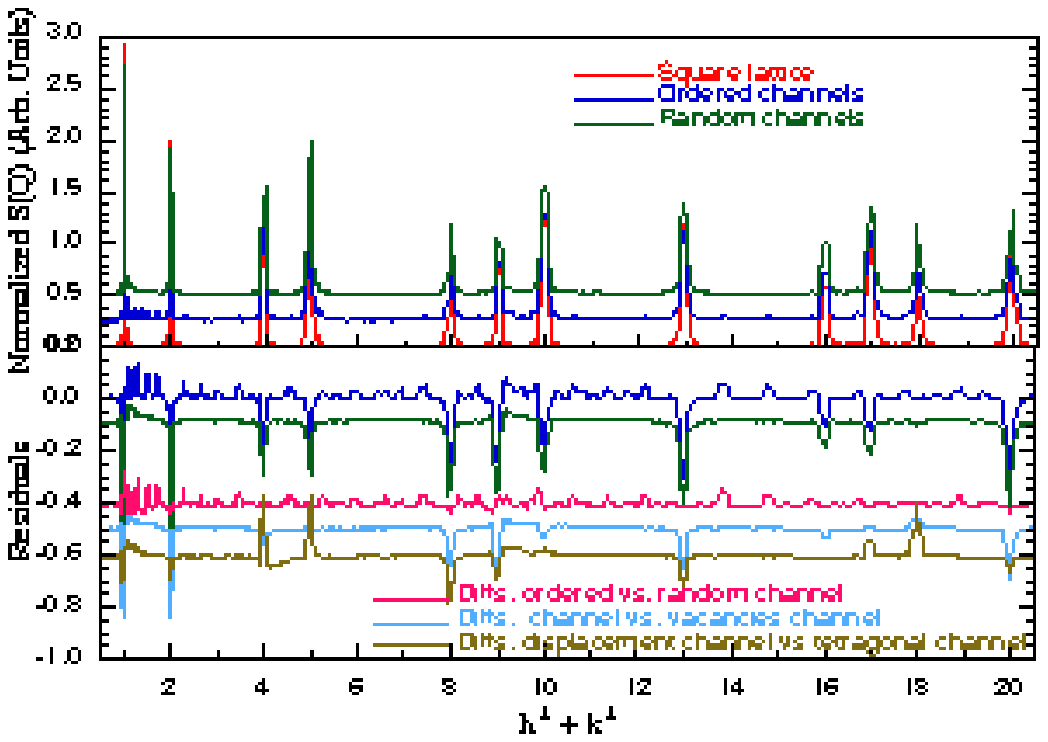}} \par}

\caption{\label{fig_channels_soq}RDF and \protect\( S(Q)\protect \) for
the lattices with tetragonal distortions forming channels. See the
comments in the caption of figure \ref{fig_vacancies_soq} about how
the lower panel has been constructed.}
\end{figure*}

\noindent \emph{Effect of networks of tetragonal distortions across
the lattice.-} The results of the calculations for these lattices
can be seen in Fig.~\ref{fig_channels_soq}. The main effect of these
distortions in the RDF is to produce many new peaks associated with
the new pair distances inside the tetragonal channels. Interestingly,
there are no noticeable differences between the periodic and aperiodic
cases. This is not the case, however, for the \( S(Q) \). The periodic
channels give rise to new superlattice diffraction peaks as happened
in all the cases where the distortions where periodic. When the periodicity
is removed, these superlattice peaks disappear and there is an enhancement
of the diffuse scattering localized around the fundamental peaks of
the square lattice, even though the RDFs for both lattices were essentially
identical. It is interesting to notice that this enhancement of the
diffuse scattering is somewhat similar to the one observed for the
channel of vacancies and channel of random displacements. However,
in this case the enhancement only occurs at the high-\( Q \) part
of the \( [i\, 0] \) reflections, with \( i \) \emph{an odd integer}.
For completeness, we have included the differences between the \( S(Q) \)
for the aperiodic channel of tetragonal distortions (Fig.~\ref{fig_channels_b})
and the ones for the channel of vacancies (Fig.~\ref{fig_vacancies_b})
and the channel of aperiodic displacements (Fig.~\ref{fig_displacements_b})
in the lower panel of Fig.~\ref{fig_channels_soq_b}. Apart from differences
in the relative intensities of the fundamental peaks, the main difference
occurs, obviously, around the high-\( Q \) part of the \( [i\, 0] \)
reflections, for the reasons commented above. The rest of the diffuse
scattering is essentially the same for the three lattices.

\subsection{A comment on the superlattice peaks}

\begin{figure*}[ht!]
{\centering \resizebox*{0.75\textwidth}{!}{\includegraphics{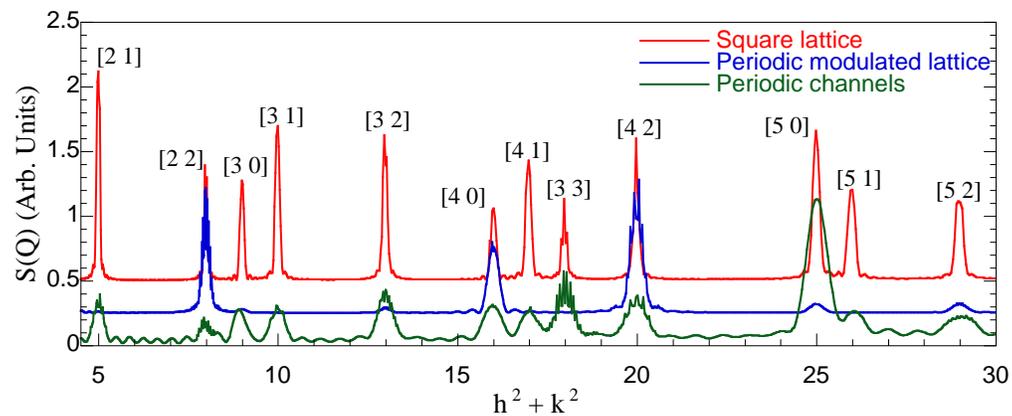}} \par}

\caption{\label{fig_basis}Structure factors for the square lattice, the periodic
modulated lattice, and the lattice with periodic tetragonal channels.
In this figure the normalization factors for the different lattices
are different, as we are only interested in the positions of the peaks
of the distorted lattices compared with the positions of the Bragg
reflections of the square lattice. }
\end{figure*}

Throughout this work, we have consistently observed that the effect
of periodic lattice distortions on the calculated structure factors
is the appearance of localized peaks, with very small intensities,
at incommensurate values of the Miller indices. We have denoted these
peaks as \emph{superlattice peaks} or \emph{superlattice reflections},
extending the common usage of these terms that are usually used for
non-stoichiometric compounds with chemical disorder\cite{Warren1990},
to the present mono-component lattices. In fact, these satellite peaks
appearing in our calculations for periodically distorted lattices
are nothing but Bragg peaks and the reason why they appear at incommensurate
positions in the previous figures is because of the \emph{naive} (and
intentional) way in which the unit cell for those lattices was chosen.
To see this, let us consider two representative examples, namely,
the periodically modulated lattice (Fig.~\ref{fig_modulated_a}) and
the lattice with periodic channels of tetragonal distortions (Fig.~\ref{fig_channels_a}).
For the modulated lattice it is easy to see that the resulting distorted
lattice can be described as another square lattice in which the smallest
motif that generates the whole lattice is another square unit cell
of lattice parameter \( 2a \) with a basis of 4 atoms. For the channel
of tetragonal distortions, the elementary motif is another square
lattice of lattice parameter \( 7a \) with a basis of 58 atoms. If
we now use the correct values for the size of the unit cell we will
obtain results as the ones depicted in Fig.~\ref{fig_basis}. As
we can see, what seemed to be satellite peaks at incommensurate positions
in \( h \) and \( k \) are nothing but Bragg reflections occurring
at the same positions as the ones of the square lattice. Of course,
the relations between the intensities of these peaks are completely
different from the ones for the simple square lattice, due to the
relative positions of the atoms inside the unit cell. Actually, these
geometrical relations lead in some cases to a cancellation of the
intensity of some peaks, what is usually called an \emph{extinction}
of the peak. This is especially noticeable in these examples for the
modulated lattice, in which all the peaks with both \( h \) and \( k \)
odd are absent, those with mixed \( h \) and \( k \) are almost
extinct, and the relative intensities of the peaks with both \( h \)
and \( k \) even are unaffected. The analysis for the lattice with
periodic channels of tetragonal distortions is more complicated. To
begin with, the original \( 40\times 40 \) lattice considered in
this work contains only \( 5\times 5 \) periodic unit cells%
\footnote{And the lattice is not truly periodic. It is interesting to notice
that the \( S(Q) \) does not reflect this fact.
}. Thus, the shapes and intensities of the peaks at very low \( Q \)
are extremely affected by the finite size of the lattice and the intensities
of these peaks are of the same order of magnitude as the oscillatory
tails of the Bragg peaks (which are also associated to the finite
size of the lattices), making it very difficult to distinguish among
them. For this reason, we repeated the calculation for a bigger lattice
containing \( 13\times 13 \) unit cells (which amounts to 9802 atoms)
and reduced the interatomic distance of the undistorted lattice to
1 Å. In doing so, the Bragg peaks are easily distinguished from the
oscillatory tails. The result of this calculation shows again that
the satellite peaks of Fig.~\ref{fig_channels_soq_b} are nothing
but Bragg peaks when a truly periodic unit cell is chosen. Furthermore,
all the reflections are present in this case. The main effect of the
complicated basis is that the intensity ratios of the reflections
are completely different from the corresponding ones for the square
lattice.

Obviously, we have not said anything new in this subsection, as all
these facts have been known for a long time\cite{Ashcroft}. However,
we think it is worthy to stress this point, as it seems that other
authors at times forget to mention these elementary facts when analyzing
diffraction patterns of lattices with periodic distortions.

\section{Conclusions}

This work is concerned with the effects of certain types of static
distortions, both periodic and aperiodic, local and collective, in
the diffraction patterns of materials. Starting from a 2D square lattice,
we have implemented different kinds of distortions to obtain new,
periodic or aperiodic, 2D lattices. The radial distribution functions
and structure factors for these lattices have been computed. The main
result coming from the calculations is that periodic distortions are
easily identified in the structure factors as new superlattice peaks
or, equivalently, as extinction of certain Bragg reflections or changes
in the relative intensities when the correct unit cell is chosen,
as is known. When the periodicity of the distortions is lost, there
are also some signatures present in the \( S(Q) \), for example,
an enhanced and asymmetric diffuse background around the fundamental
peaks of the original lattice or oscillations of this same background.
However, it is very difficult, if not impossible, to assess the exact
nature of these distortions (i.e. whether they are vacancies or random
displacements, for example). This fact is nothing but the realization,
for the particular examples presented in this work, of the aforementioned
uniqueness problem in the determination of the real structure of a
material taking as a starting point the diffraction pattern.

It is important to stress at this point that the experimental situation
can be even more misleading than the results presented in this work,
as experimental measurements are always subject to additional effects
we have not taken into account in our simulations. These include limited
\( Q \) resolution, instrumental broadening, and noise. Instrumental
broadening, especially, can very easily mask some of the peak splittings
we obtain in our calculations. Regarding the calculated radial distribution
functions, in some cases we have seen that they provide more intuitive
information than the \( S(Q) \) for the kinds of distortions considered
in this work. This fact strongly supports the idea that PDF measurements
can be crucially important in making progress in the problem of the
determination of the structure of complex systems. However, this approach
is not exempt to limitations. On one hand, it is necessary to use
very intense sources in order to be able to measure the diffraction
pattern up to very large values of \( Q \) (\textasciitilde{}35--45
Å\( ^{-1} \) typically). Fortunately, with the availability of very
intense synchroton and neutron sources this problem is partially alleviated,
and these wave-vectors are experimentally achievable. On the other
hand, the main limitation in extracting information from experimentally
(Fourier-transformed diffraction data) determined pair distribution
functions is the effect of thermal broadening that can very easily
mask some of the features associated to local distortions. We have
seen some examples of this problem when studying the effects of small
domains with tetragonal distortions in the radial distribution function:
the new peaks associated with the new, shorter distances of the tetragonal
phase are closer to the peaks associated to the pair distances of
the square lattice than the width of those peaks, making it very difficult
to distinguish them. What is even worse, the radial distribution function
is completely oblivious to certain kinds of distortions (as it happened
with the vacancies configurations). It is in these situations where
X-ray absorption fine structure could play an important role, as this
type of probe is sensitive to local distortions and to the different
species present in the sample. The main limitation in this case is
associated to the fact that the XAFS spectra contain multiple scattering
contributions and, thus, the kinematic limit is not applicable to
the analysis of the data, making it extremely difficult to extract
information about the radial distribution at long distances from the
probed atom.

We also want to emphasize that, obviously, we are aware of the fact
that the kind of distortions we have studied in this work hardly exist
in real materials. They are simple illustrative examples of how easy
it is to miss important structural information about local and collective
distortions if we blindly trust structural models obtained by only
applying conventional diffraction analysis methods. Our intention
is to motivate awareness about this potential problem. The emerging
picture of complex solids strongly supports the idea that local and
nanoscale distortions exist (probably even more complicated ones than
the toy models we have studied in this work), there are mechanisms
that make them stable, and they play an important role in the properties
of these systems. However, as has been shown throughout this work,
the signatures of such aperiodic distortions are subtle and misleading
interpretations can be reached if special care is not paid to their
analysis.

We think that the above paragraphs serve to further stress the point
already made in the Introduction: It is necessary to use a combination
of experimental methods that probe all the relevant length scales
(x-ray and neutron diffraction experiments, XAFS, PDF, etc). At the
same time, a combination of structural analysis methods (PDF analysis,
reverse Monte Carlo simulations, etc), beyond the standard Rietveld
analysis, is necessary in order to reach a consistent picture of the
structure of complex solids that can lead us to understand and exploit
their properties.

\begin{acknowledgments}
This work was supported by DOE DP and OBES Division of Chemical Sciences
under Contract W-7405
\end{acknowledgments}

\end{document}